\documentclass[12pt]{extarticle}
\usepackage{setspace}

\usepackage[T1]{fontenc}

\usepackage[latin1]{inputenc}
\usepackage{epsfig}
\usepackage[english,french]{babel}
\usepackage{color}
\usepackage{graphicx}

\usepackage{dcolumn}

\usepackage{moreverb}

\usepackage{amsmath,amssymb,amsfonts}

\begin{document}
\numberwithin{equation}{section}
\newcommand{\boxedeqn}[1]{%
  \[\fbox{%
      \addtolength{\linewidth}{-2\fboxsep}%
      \addtolength{\linewidth}{-2\fboxrule}%
      \begin{minipage}{\linewidth}%
      \begin{equation}#1\end{equation}%
      \end{minipage}%
    }\]%
}

%\boxedeqn{}

\newsavebox{\fmbox}
\newenvironment{fmpage}[1]
     {\begin{lrbox}{\fmbox}\begin{minipage}{#1}}
     {\end{minipage}\end{lrbox}\fbox{\usebox{\fmbox}}}

\raggedbottom
\onecolumn
%\pagestyle{headings}

%\begin{flushleft}
\begin{center}
\title*{{\LARGE{\textbf{Construction of classical superintegrable systems with higher order integrals of motion from ladder operators}}}}
\end{center}
Ian Marquette
\newline
Department of Mathematics, University of York, Heslington, York, UK. YO10 5DD
\newline
im553@york.ac.uk
\newline
\newline
We construct integrals of motion for multidimensional classical systems from ladder operators of one-dimensional systems. This method can be used to obtain new systems with higher order integrals. We show how these integrals generate a polynomial Poisson algebra. We consider a one-dimensional system with third order ladders operators and found a family of superintegrable systems with higher order integrals of motion. We obtain also the polynomial algebra generated by these integrals. We calculate numerically the trajectories and show that all bounded trajectories are closed. 
\newline
\section{Introduction}
Over the years, many articles were devoted to superintegrability [1-15]. For a review of superintegrability of two-dimensional systems we refer the reader to the Ref.14. The relation between constants of motion and ladder operators in classical and quantum mechanics was acknowledged by several authors [3,5,6,16-24]. Detailed discussions of the relation between integrals and ladder operators for the two-dimensional harmonic oscillator, anisotropic harmonic oscillator and Kepler-Coulomb systems were done [17,18]. Ladder operators are more used in context of quantum mechanics. They can provide the wave functions and the energy spectrum of the corresponding Schrodinger equation and the eigenstates of the annihilation operator are related to coherent states [25]. In quantum mechanics, these raising and lowering operators are also related to supercharges and supersymmetric quantum mechanics. The quantum superintegrable systems with third order integrals of motion [15,16] were related to supersymmetric quantum mechanics (SUSYQM) [26,27] and higher order supersymmetric quantum mechanics (HSQM)[28,29,30]. A method to generate quantum superintegrable systems from supersymmetry was presented in Ref.31. This method allows to generate systems with higher order integrals of motion. In a recent article, we discussed how ladder operators can be used to generate higher order integrals of motion and superintegrable systems in context of quantum mechanics [32]. We imposed the separation of variables in Cartesian coordinates and the order of the ladder operators were arbitrary. These relations between quantum superintegrable systems, integrals of motion, polynomial algebras, ladders operators and supersymmetry are interesting and provide new insight. In the light of these results the study of systems with ladder operators appears to be important also in regard of superintegrable systems. The classification of systems with first or second order ladder operators in $E_{2}$ was discussed [33]. Systems with third [16,29] and also fourth [30] order ladder operators were discussed in context of supersymmetric quantum mechanics. 
\newline
The purpose of this paper is to discuss how the method developped in context of quantum mechanics to obtain integrals of motion and polynomial algebras from ladder operators [32] can be applied in classical mechanics. The method allows to obtain multidimensional superintegrable systems however we will focus on two-dimensional systems. We will also discuss systems with third order ladder operators. To our knowledge, the study of classical systems with higher order ladder operators is also an unexplored subject. We will point out that ladder operators appear important in regard of classical superintegrable systems.
\newline
Let us present the organization of this paper. In Section 2, we will present a method to generate higher order integrals of motion and new classical superintegrable systems from one-dimensional systems with ladder operators. We present the general polynomial Poisson algebra obtained from these integrals of motion. In Section 3, we consider a system that we studied in an earlier article concerning classical superintegrable systems in two-dimensional Euclidean space separable in Cartesian coordinates with a second and a third order integrals [17]. We show that this system possesses third order ladder operators. We use these operators and results of Section 2 to generate new superintegrable systems. We present their integrals of motion and polynomial Poisson algebras. We obtain their trajectories. They are deformed Lissajous' figure [34]. We show also that all bounded trajectories are closed [35]. 
\newline
Before proceeding with the results let us recall a few important definitions. In classical mechanics a Hamiltonian system with Hamiltonian H and integrals of motion $X_{a}$
\begin{equation}
H=\frac{1}{2}g_{ik}p_{i}p_{k}+V(\vec{x},\vec{p}),\quad X_{a}=f_{a}(\vec{x},\vec{p}),\quad a=1,..., n-1 \quad,
\end{equation}
is called completely integrable (or Liouville integrable) if it
allows n integrals of motion (including the Hamiltonian) that are
well defined functions on phase space, are in involution
$\{H,X_{a}\}_{p}=0$, $\{X_{a},X_{b}\}_{p}=0$, a,b=1,...,n-1 and
are functionally independent ($\{,\}_{p}$ is a Poisson bracket).
A system is superintegrable if it is integrable and allows further
integrals of motion $Y_{b}(\vec{x},\vec{p})$, $\{H,Y_{b}\}_{p}=0$,
b=n,n+1,...,n+k that are also well defined functions on phase
space and the integrals $\{H,X_{1},...,X_{n-1},Y_{n},...,Y_{n+k}\}$
are functionally independent. A system is maximally superintegrable if the set contains 2n-1 functions. The integrals
$Y_{b}$ are not required to be in evolution with $X_{1}$,...$X_{n-1}$, nor with each other. 
\section{Ladder operators and integrals of motion}
Let us consider a classical two-dimensional Hamiltonian separable in Cartesian coordinates
\begin{equation}
H(x_{1},x_{2},P_{1},P_{2})=H_{1}(x_{1},P_{1})+H_{2}(x_{2},P_{2}),
\end{equation}
for which polynomial ladder operators ($A_{x_{i}}$ and $A_{x_{i}}^{+}$) exist. These operators satisfy the relations
\begin{equation}
\{H_{i},A_{x_{i}}^{+}\}_{p}=\lambda_{x_{i}}A_{x_{i}}^{+},\quad \{H_{i},A_{x_{i}}^{-}\}_{p}=-\lambda_{x_{i}}A_{x_{i}}^{-}\quad ,
\end{equation}
\begin{equation}
\{A_{x_{i}}^{-},A_{x_{i}}^{+}\}_{p}=P_{i}(H_{i}),\quad A_{x_{i}}^{-}A_{x_{i}}^{+}=A_{x_{i}}^{+}A_{x_{i}}^{-}=Q_{i}(H_{i}),\quad i=1,2
\end{equation}
where $P_{i}(H_{i})$ and $Q_{i}(H_{i})$ are polynomials. These relations are the classical analog of relation imposed in Ref.32. They are satisfy for many well known superintegrable systems such the Harmonic oscillator and the Smorodinsky-Winternitz potentials that allow separation of variables in Cartesian coordinates. From the relation (2.2) the operators $f_{1}=A_{x_{1}}^{+m_{1}}A_{x_{2}}^{-m_{2}}$ and $f_{2}=A_{x_{1}}^{-m_{1}}A_{x_{2}}^{+m_{2}}$ Poisson commute with the Hamiltonian H given by Eq.(2.1) if $m_{1}\lambda_{x_{1}}-m_{2}\lambda_{x_{2}}=0$ with $m_{1}$,$m_{2}$ $\in \mathbb{Z}^{+}$. The following sums are also polynomial integrals of the Hamiltonian H
\begin{equation}
I_{1}=A_{x_{1}}^{+ m_{1}}A_{x_{2}}^{- m_{2}}- A_{x_{1}}^{- m_{1}}A_{x_{2}}^{+ m_{2}}, \quad I_{2}=A_{x_{1}}^{+ m_{1}}A_{x_{2}}^{- m_{2}}+ A_{x_{1}}^{- m_{1}}A_{x_{2}}^{+ m_{2}}\quad .
\end{equation}
By construction the Hamiltonian has the following second order integral from separation of variables.
\begin{equation}
K=H_{1}-H_{2}.
\end{equation}
The Hamiltonian H given by Eq.(2.1) is thus superintegrable. 
\newline
We will now interested by the algebraic structure generated by these integrals. In quantum mechanics quadratic [36,37], cubic [14-16] and higher order polynomial algebras [32] can be written as deformed oscillator algebras [38]. The Fock type unitary representations can be used to obtain the energy spectrum. This is the classical equivalent of the polynomial algebra obtained in Ref. 24. The Eq.(2.2) and (2.3) are classical analog of deformed oscillator algebra [38]. Such algebras were discussed by A.V.Tsiganov in Ref. 39. The Eq(2.3) can also be interpreted as classical analog of the factorization method in supersymmetric quantum mechanics [40]. We construct from integrals given by Eq.(2.4) and (2.5) the following polynomial Poisson algebra
\begin{equation}
\{K,I_{1}\}_{p}=2\lambda I_{2},\quad \{K,I_{2}\}_{p}=2\lambda I_{1},\quad \{I_{1},I_{2}\}_{p}=2 Q_{1}(\frac{1}{2}(H+K))^{m_{1}-1}
\end{equation}
\[Q_{2}(\frac{1}{2}(H-K))^{m_{2}-1}[m_{2}^{2}Q_{1}(\frac{1}{2}(H+K))P_{2}(\frac{1}{2}(H-K))-m_{1}^{2}Q_{2}(\frac{1}{2}(H-K))P_{1}(\frac{1}{2}(H+K))]. \]
This is the classical analog of the algebra obtained in Ref. 32. We can relate the polynomial by the following relations. The polynomial algebra of superintegrable systems in classical mechanics play an important role in their classification [41].
\newline
The method can also be extended in N-dimensions by forming the following integrals
\begin{equation}
I_{ij}=A_{x_{i}}^{+ m_{i}}A_{x_{j}}^{- m_{j}}-A_{x_{i}}^{- m_{i}}A_{x_{j}}^{+ m_{j}},\quad J_{ij}=A_{x_{i}}^{+ m_{i}}A_{x_{j}}^{- m_{j}}+A_{x_{i}}^{- m_{i}}A_{x_{j}}^{+ m_{j}} ,
\end{equation}
\[ K_{ij}=H_{x_{i}}-H_{x_{j}},\quad 0\leq i < j \leq N.\]
\section{Construction of new superintegrable systems}
Let us present a system obtained in Ref. 12 and studied in Ref. 13.
\begin{equation}
H = \frac{P_{1}^{2}}{2} + \frac{P_{2}^{2}}{2} +
\frac{\omega^{2}}{2}x_{2}^{2} + V(x_{1}).
\end{equation}
where the potential $V(x_{1})$ satisfies a quartic equation
\begin{equation}
-9V(x_{1})^{4} +
14\omega^{2}x_{1}^{2}V(x_{1})^{3}+(6d-15\frac{\omega^{4}}{2}x_{1}^{4})V(x_{1})^{2}+(\frac{3\omega^{6}}{2}x_{1}^{6}-2d\omega^{2}x_{1}^{2})V(x_{1})
\end{equation}
\[+(cx_{1}^{2}-d^{2}-d\frac{\omega^{4}}{2}x_{1}^{4}-\frac{\omega^{8}}{16}x_{1}^{8})=0.\]
This Hamiltonian has two integrals 
\begin{equation}
A = \frac{P_{1}^{2}}{2} -
\frac{P_{2}^{2}}{2} - \frac{\omega^{2}}{2}x_{2}^{2} + V(x_{1}),
\end{equation}
\[ B = -x_{2}P_{1}^{3} + x_{1}P_{1}^{2}P_{2} + (\frac{\omega^{2}}{2}x_{1}^{2} - 3V(x_{1}))x_{2}P_{1}-\frac{1}{\omega^{2}}(\frac{\omega^{2}}{2}x_{1}^{2}-3V(x_{1}))V_{x_{1}}(x_{1})P_{2}.\]
In the quantum case V satifies a fourth order differential
equation [16]
\begin{equation}
\hbar^{2}V^{(4)}(x_{1})=12\omega^{2}x_{1}V'(x_{1})+6(V^{2}(x_{1}))''-2\omega^{2}x_{1}^{2}V''(x_{1})+2\omega^{4}x_{1}^{2}
\end{equation}
that can be solved in terms of the fourth Painlev\'e transcendent [42]. In general,
Eq.(3.2) has 4 roots and the expressions for them are quite
complicated. A special case occurs if $\omega^{2}$,c and d satisfy $c=\frac{2^{3}\omega^{8}b^{3}}{3^{6}}$ and $d=\frac{\omega^{4}b^{2}}{3^{3}}$ where b is an arbitrary constant. Then  Eq.(3.2) has a double root
and we obtain
\begin{equation}
V(x_{1})=\frac{\omega^{2}}{18}(2b + 5x_{1}^{2} + \epsilon 4x_{1}\sqrt{b+x_{1}^{2}})
\end{equation}
The potentials $V(x_{1})$  is a deformed harmonic oscillators. The Hamiltonian given by Eq.(3.1) with potential given by Eq.(3.5) reduces to isotropic harmonic oscillator or anisotropic harmonic oscillator (with ratio 3:1) when $b=0$. For $V(x_{1})$ satisfying Eq.(3.5) the cubic algebra is
\begin{equation}
\{A,B\}_{p}=C,\quad \{A,C\}_{p}=-4\omega^{2}B,\quad \{B,C\}_{p}=8A^{3} + 12HA^{2}
\end{equation}
\[ - 4H^{3} -4\frac{4b^{2}\omega^{4}}{27}A + \frac{4b^{3}\omega^{6}}{729}.\]
\subsection{Deformed isotropic and anisotropic harmonic oscillator}
Let us consider the following one-dimensional systems with potential given by Eq.(3.5)
\begin{equation}
H_{1}=\frac{P_{1}^{2}}{2}+\frac{\omega^{2}}{18}(2b_{1} + 5x_{1}^{2} + \epsilon_{1} 4x_{1}\sqrt{b_{1}+x_{1}^{2}}).
\end{equation}.
Like its quantum equivalent, this system has third order creation and annihilation operators 
\begin{equation}
a_{x_{1}}^{+}=P_{1}^{3}-i\omega x_{1}P_{1}^{2}+(\frac{b_{1}^{2}\omega^{2}}{3}+ \frac{\omega^{2}x_{1}^{2}}{3}+\frac{2}{3}\epsilon_{1}\omega^{2}x_{1}\sqrt{b_{1}^{2}+x_{1}^{2}})P_{1}
\end{equation}
\[ +i(-\frac{1}{3}b_{1}^{2}\omega^{3}x_{1}-\frac{13}{27}\omega^{3}x_{1}^{3}-\frac{2}{27}b_{1}^{2}\epsilon_{1}\sqrt{b_{1}^{2}+x_{1}^{2}}-\frac{14}{27}\epsilon_{1}\omega^{3}x_{1}^{2}\sqrt{b_{1}^{2}+x_{1}^{2}})
\]
\begin{equation}
a_{x_{1}}^{-}=P_{1}^{3}+i\omega x_{1}P_{1}^{2}+(\frac{b_{1}^{2}\omega^{2}}{3}+ \frac{\omega^{2}x_{1}^{2}}{3}+\frac{2}{3}\epsilon_{1}\omega^{2}x_{1}\sqrt{b_{1}^{2}+x_{1}^{2}})P_{1}
\end{equation}
\[ -i(-\frac{1}{3}b_{1}^{2}\omega^{3}x_{1}-\frac{13}{27}\omega^{3}x_{1}^{3}-\frac{2}{27}b_{1}^{2}\epsilon_{1}\sqrt{b_{1}^{2}+x_{1}^{2}}-\frac{14}{27}\epsilon_{1}\omega^{3}x_{1}^{2}\sqrt{b_{1}^{2}+x_{1}^{2}}).
\]
satifying
\begin{equation}
\{H,a_{x_{1}}^{+}\}_{p}=i\omega a_{x_{1}}^{+},\quad \{H,a_{x_{1}}^{-}\}_{p}=-i\omega a_{x_{1}}^{-}.
\end{equation}
The results of Section 2 allows us to form the following N-dimensional classical superintegrable system
\begin{equation}
H=\sum_{j=1}^{N}\frac{P_{j}^{2}}{2}+\frac{\omega^{2} k_{j}^{2}}{18}(2b_{j} + 5x_{j}^{2} + \epsilon_{j} 4x_{j}\sqrt{b_{j}+x_{j}^{2}}).
\end{equation}
It has ladder operators of the same form given by Eq.(3.8) and (3.9) in each axis. The polynomial $P_{j}(H_{j})$ and $Q_{j}(H_{j})$ are given by
\begin{equation}
P_{j}(H_{j})=24i\omega k_{j}H_{j}^{2}+\frac{16}{3}i(-b_{j}\omega^{3}k_{j}^{3}+b_{j}^{2}\omega^{3}k_{j}^{3})H_{j}+\frac{2}{27}i(4b_{j}^{2}\omega^{5}k_{j}^{5}-8b_{j}^{3}\omega^{5}k_{j}^{5}+3b_{j}^{4}\omega^{5}k_{j}^{5})
\end{equation}
\begin{equation}
Q_{j}(H_{j})=\frac{2}{729}(18H_{j}+(b_{j}-2)b_{j}\omega_{j}^{2})^{2}(9H_{j}+b_{j}(2b_{j}-1)\omega_{j}^{2})
\end{equation}
For the case N=2 the integrals are thus given by Eq.(2.4) and (2.5). These integrals generate the polynomial Poisson algebra given by Eq.(2.7) with $P_{1}(H_{1})$,$P_{2}(H_{2})$, $Q_{1}(H_{1})$ and $Q_{2}(H_{2})$ given by Eq.(3.12) and (3.13). The integrals $I_{1}$ and $I_{2}$ are polynomials in the momenta of order $3^{m_{1}+m_{2}}-1$ and $3^{m_{1}+m_{2}}$ respectively. The order of the polynomial algebra is $2\times 3^{m_{1}+m_{2}-1}$. In the N-dimensional case the integrals are given by Eq.(2.7). Only 2N-1 of these integrals are functionally independent and this system is maximally superintegrable. The integrals $I_{ij}$, $J_{ij}$ and $K_{ij}$ are polynomials in the momenta of order $3^{m_{i}+m_{j}}-1$, $3^{m_{i}+m_{j}}$ and 2 respectively.
\newline
The trajectories are obtained numerically directly from the equations of motion. We present trajectories for the cases N=2 and N=3 for specific parameters (Fig1, Fig2, Fig3 and Fig4). The bounded trajectories are closed and correspond to deformed Lissajous's figures.
\newline
\begin{figure}
 \begin{minipage}[b]{.23\linewidth}
  \centering\epsfig{figure=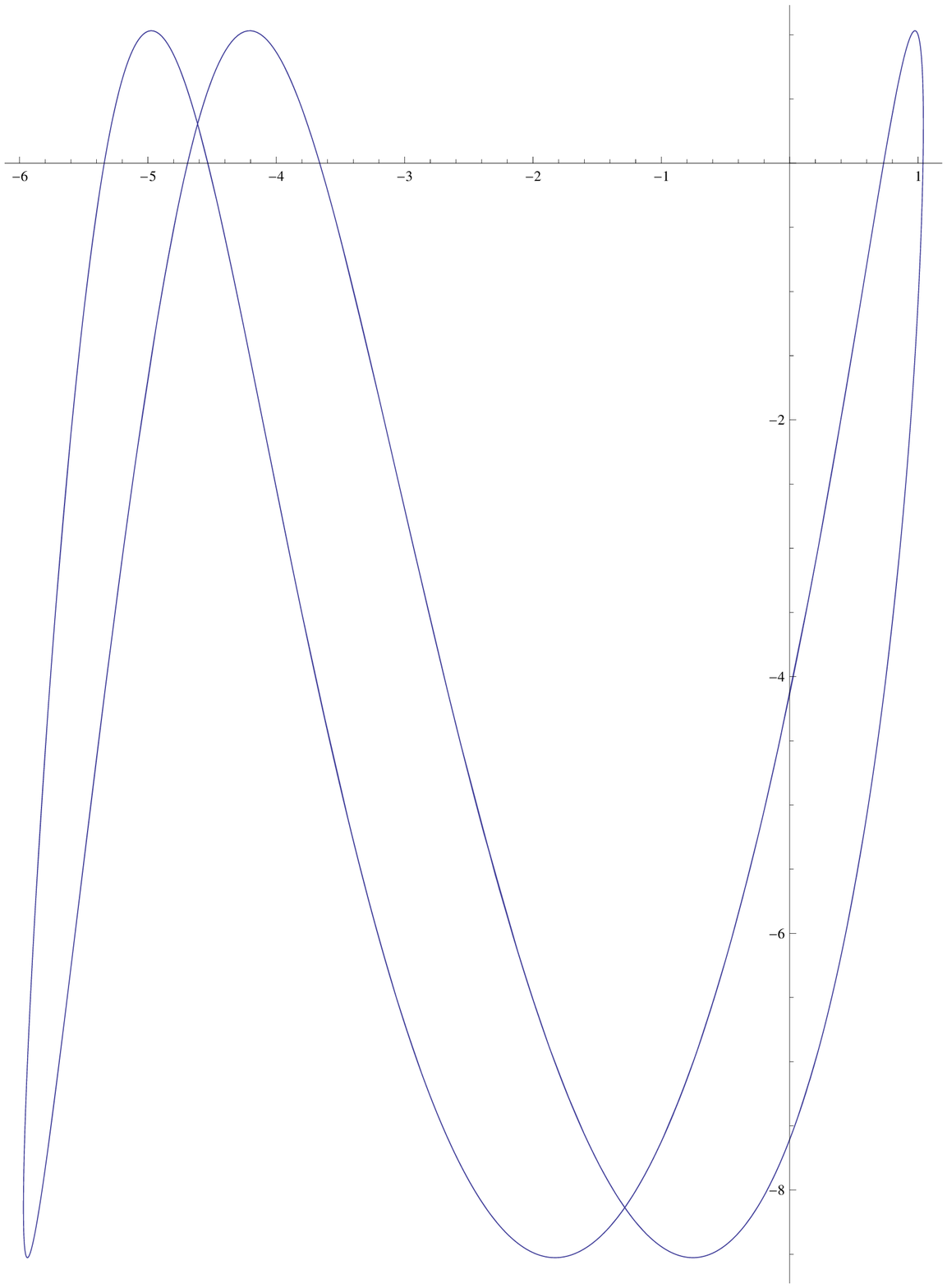,width=\linewidth}
% \unnumberedcaption{Fig. 1}
\caption{} 
 \end{minipage} \hfill
 \begin{minipage}[b]{.23\linewidth}
  \centering\epsfig{figure=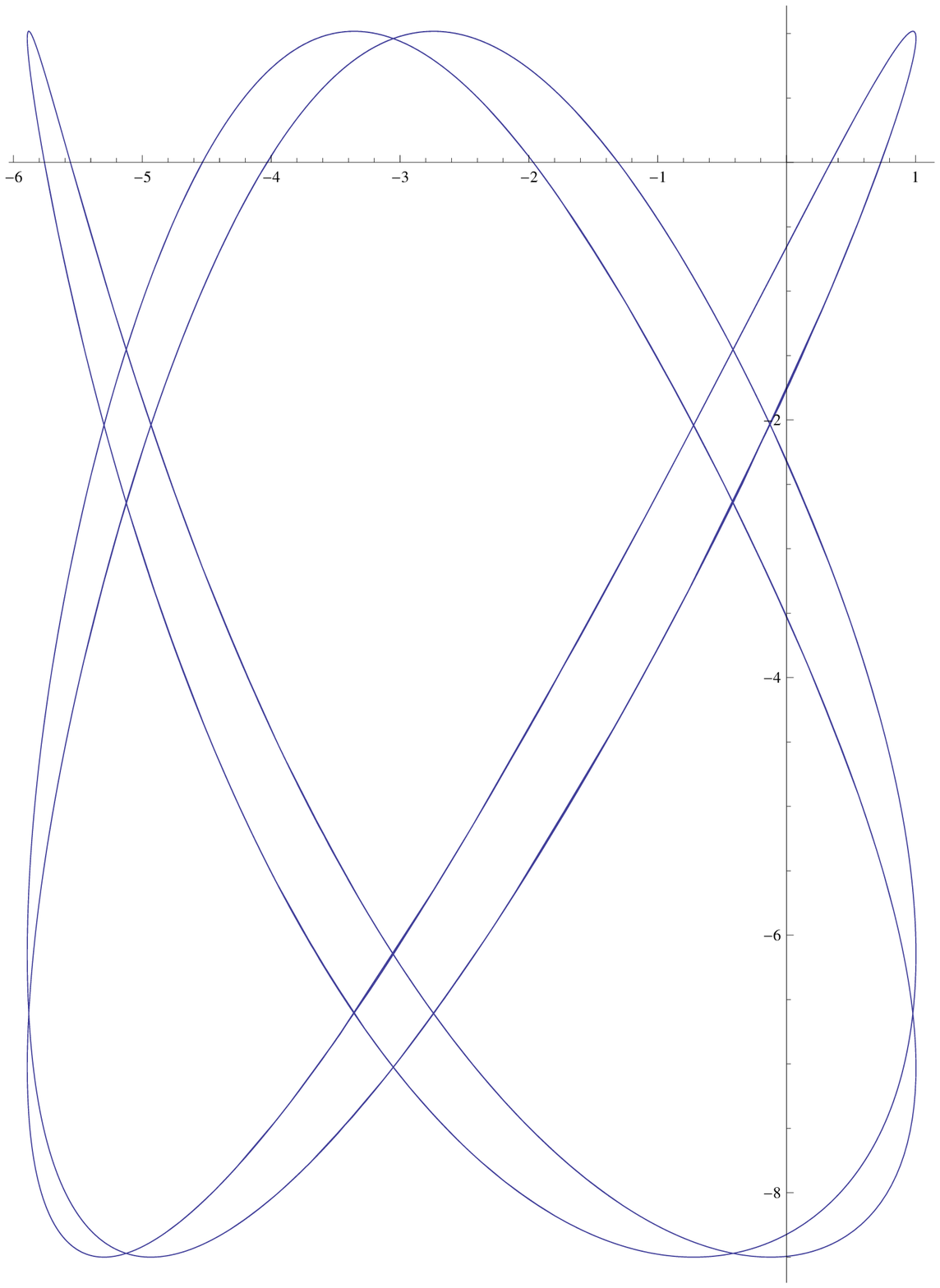,width=\linewidth}
%  \unnumberedcaption{Fig. 2}
\caption{}
 \end{minipage}\hfill
 \begin{minipage}[b]{.23\linewidth}
  \centering\epsfig{figure=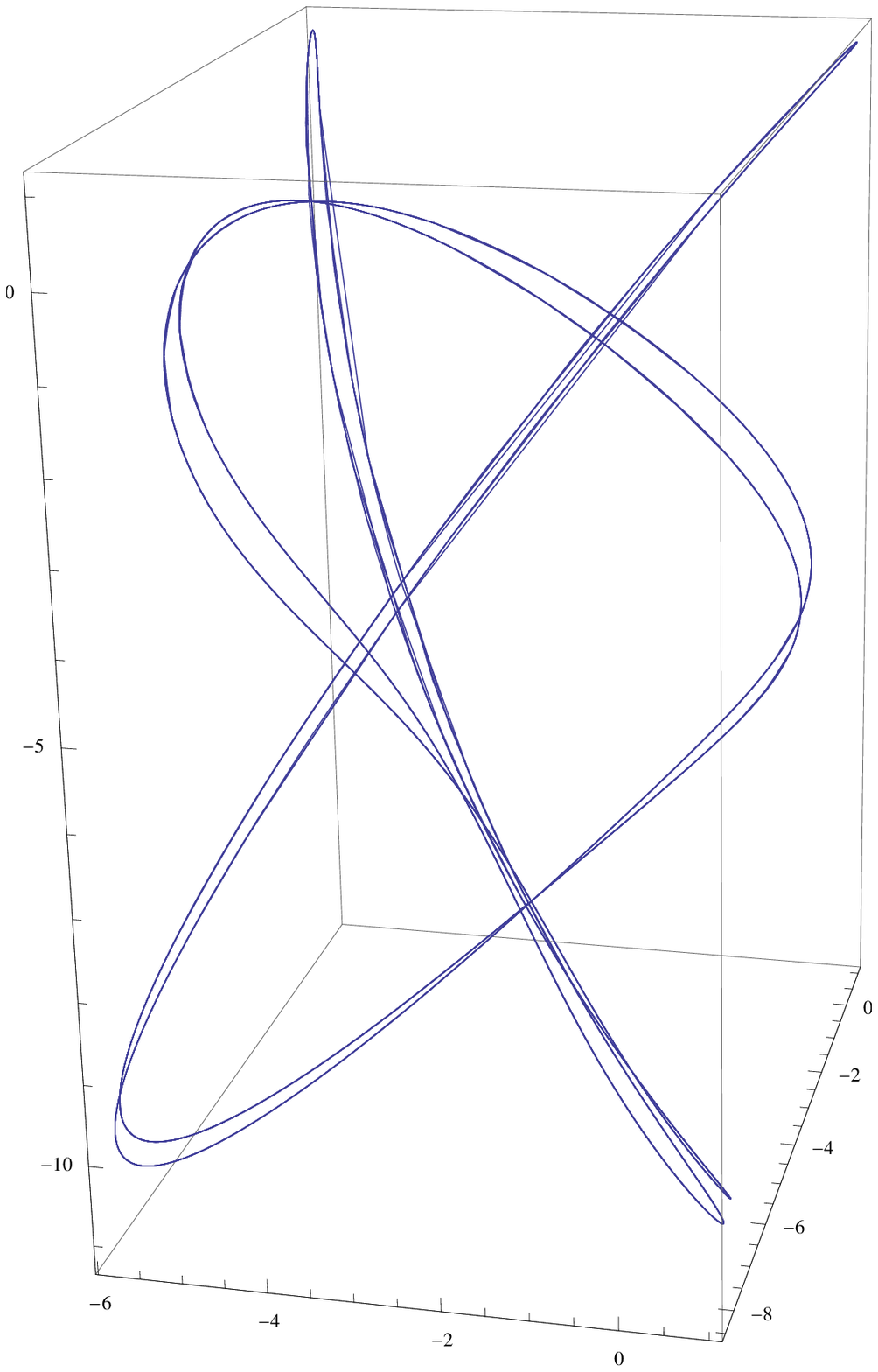,width=\linewidth}
%  \unnumberedcaption{Fig. 3}
\caption{}
 \end{minipage} \hfill
 \begin{minipage}[b]{.23\linewidth}
  \centering\epsfig{figure=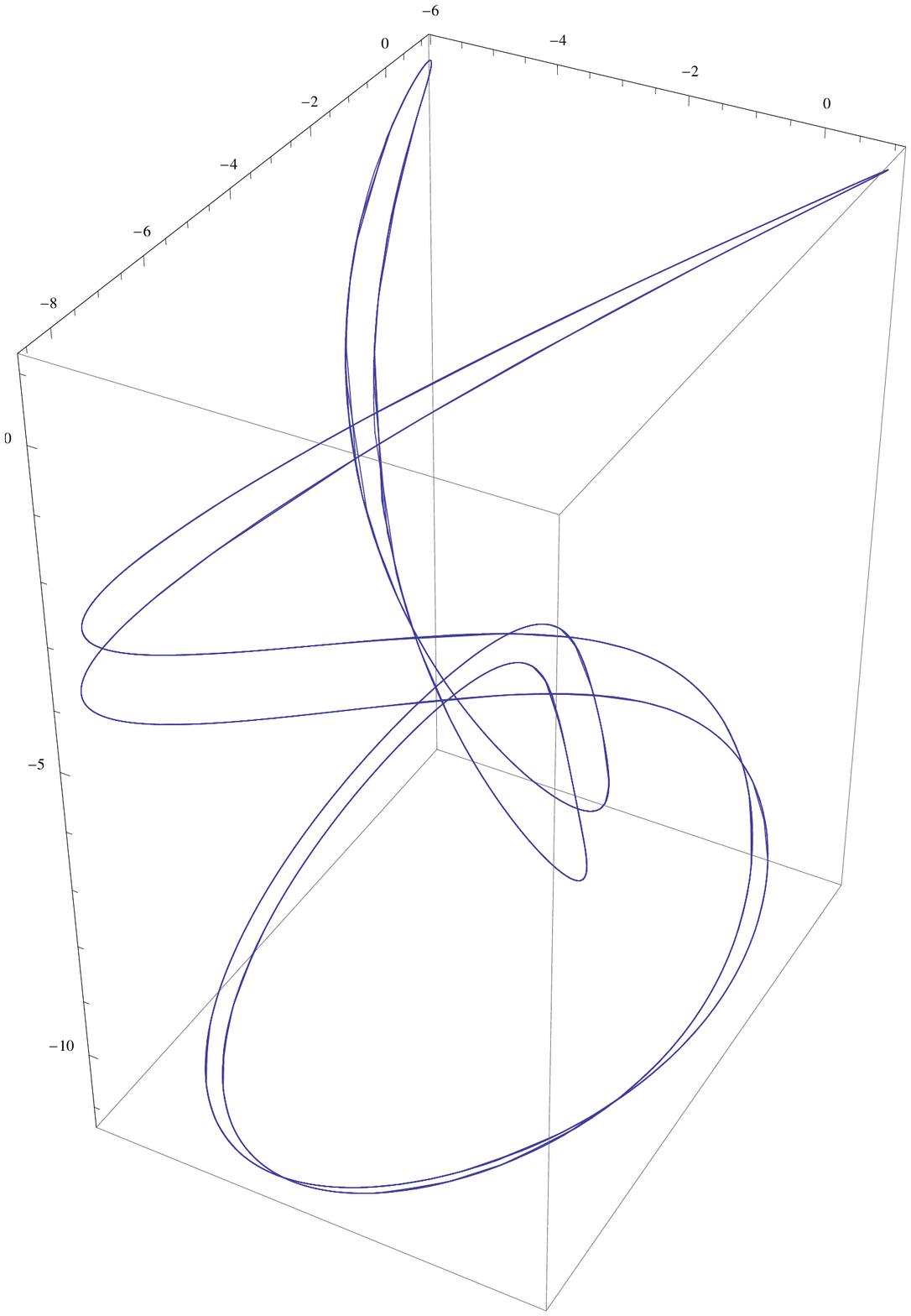,width=\linewidth}
%  \unnumberedcaption{Fig. 4}
\caption{}
 \end{minipage}
\end{figure}
\section{Conclusion}
Other methods to obtain superintegrable systems were discussed recently: the method of coupling constant metamorphosis in context of higher order integrals of motion [43] and the method of symmetry reduction [44,45]. In this paper, we constructed for a class of classical systems integrals of motion from ladder operators. This method allows to generate new classical superintegrable systems with higher order integrals of motion from one-dimensional Hamiltonian for which we know the ladder operators. From the requirement that ladder operators satisfy the classical analog of deformed oscillator algebras we obtained the polynomial Poisson algebra for the superintegrable systems.
\newline
We considered the one-dimensional system given by Eq.(3.5) for which the potential satisfy a quartic equation. This system has third order ladder operators. From this systems and the ladder operators and the method of Section 2, we constructed the integrals of motion and polynomial Poisson algebra for a new two-dimensional suprintegrable systems. A family of superintegrable systems in N dimensions can be generated. We present also the trajectories in the two and three dimensional case and obtain deformed Lissajous figures. All bounded trajectories are closed.
\newline
\newline
We discuss in this paper systems with third order ladder operators. Classical systems with higher order ladder operators were not systematically studied. The study of such systems could provide new superintegrable systems with higher integral of motion. 
\newline
\textbf{Acknowledgments} The research of I.M. was supported by a postdoctoral research fellowship from FQRNT of Quebec. The author thanks P.Winternitz for very helpful comments and discussions.

\section{\textbf{References}}

1. V.Fock, Z.Phys. 98, 145-154 (1935).
\newline
2. V.Bargmann, Z.Phys. 99, 576-582 (1936).
\newline
3. J.M.Jauch and E.L.Hill, Phys.Rev. 57, 641-645 (1940).
\newline
4. J.Fris, V.Mandrosov, Ya.A.Smorodinsky, M.Uhlir and P.Winternitz,  Phys.Lett. 16, 354-356 (1965).
\newline
5. P.Winternitz, Ya.A.Smorodinsky, M.Uhlir and I.Fris, Yad.Fiz. 4,
625-635 (1966). (English translation in Sov. J.Nucl.Phys. 4,
444-450 (1967)).
\newline
6. N.W.Evans, Phys.Rev. A41, 5666-5676 (1990), J.Math.Phys. 32,
3369-3375 (1991).
\newline
8. E.G.Kalnins, W.Miller Jr and S.Post, J.Phys.A: Math Theor. 40,  11525-11538 (2007).
\newline
9. P.Winternitz anf I.Yurdusen, J.Math.Phys. 47, 103509 (2006).
\newline
10. J.Berube and P.Winternitz, J.Math.Phys. 45 (5) 1959-1973 (2004).
\newline
11. S.Gravel and P.Winternitz, J.Math.Phys. 43(12), 5902 (2002).
\newline
12. S.Gravel, J.Math.Phys. 45(3), 1003-1019 (2004).
\newline
13. I.Marquette and P.Winternitz, J.Math.Phys. 48(1) 012902
(2007).
\newline
14. I.Marquette and P.Winternitz, J. Phys. A: Math. Theor. 41, 304031 (2008).
\newline
15. I.Marquette, J. Math. Phys. 50, 012101 (2009).
\newline
16. I.Marquette, J. Math. Phys. 50 095202 (2009).
\newline
17. V.A.Dulock and H.V.McIntosh, Am. J.Phys. 33, 109 (1965).
\newline
18. A.Cisneros and H.V.McIntosh, J.Math.Phys. 11 3 870 (1970).
\newline
19. R.D.Mota, V.D.Granados, A.Queijeiro and J.Garcia, J.Math A: Math. Gen. 35 2979-2984 (2002).
\newline
20. J.M.Lyman and P.K.Aravind, J.Phys. A: Math. Gen. 26, 3307-3311 (1993).
\newline
21. R.D.Mota, J.Garcia and V.D.Granados, J.Math A: Math. Gen. 34 2041-2049 (2001).
\newline
22. Y.F. Liu, W.J.Huo and J.Y.Zeng, Phys. Rev. A 58 2 862-868 (1998)
\newline
23. Mikhail Plyushchay, Int.J.Mod.Phys.A15:3679-3698,2000. 
\newline
24. Francisco Correa, Vit Jakubsky, Mikhail S. Plyushchay, Annals Phys.324:1078-1094,2009,.
\newline
25. D.J.Fern\'andez, V.Hussin et L.M.Nieto, J.Phys. A 27 3547-3564 (1994).
\newline
26. E.Witten, Nucl.Phys. B188, 513 (1981); E.Witten, Nucl.Phys. B202, 253-316 (1982)
\newline
27. G.Junker, Supersymmetric Methods in Quantum and Statistical Physics, Springer, New York, (1995).
\newline
28. A.Andrianov, M.Ioffe and V.P.Spiridonov, Phys.Lett. A174, 273 (1993). 
\newline
29. A.Andrianov, F.Cannata, M.Ioffe and D.Nishnianidze, Phys.Lett.A, 266,341-349 (2000).
\newline
30. J.M.Carballo, D.J.Fernandez C, J.Negro and L.M.Nieto, J.Phys. A: Math. Gen. 37, 10349-10362 (2004).
\newline
31. I.Marquette, J.Math.Phys. 50 122102 (2009).
\newline
32. I.Marquette, J.Phys A: Math. Gen. 43 135203 (2010).
\newline
33. C.P.Boyer and W.Miller Jr., J.Math.Phys. 15, 9 (1974).
\newline
34. J.Lissajous, Ann. Chim. Phys. 51, 147-231 (1857).
\newline
35. N.N.Nekhoroshev,Action-angle variables and their
generalizations.Trudy Mskov.Mat.Obshch.26,181-198 (1972).
\newline
36. D.Bonatsos, C.Daskaloyannis and K.Kokkotas, Phys.Rev. A48,
R3407-R3410 (1993).
\newline
37. D.Bonatsos, C.Daskaloyannis and K.Kokkotas, Phys. Rev.A 50, 3700-3709 (1994), C.Daskaloyannis, J.Math.Phys. 42, 1100 (2001).
\newline
38. C.Daskaloyannis, Generalized deformed oscillator and nonlinear
algebras, J.Phys.A: Math.Gen 24, L789-L794 (1991).
\newline
39. A.V. Tsiganov, J.Phys. A, Math. Gen 33 41 7407-7422 (2000).
\newline
40. S.Kuru and J.Negro, Ann. Phys. 323 2 413-431 (2008).
\newline
41 C.Daskaloyannis and K.Ypsilantis, J.Math.Phys. 47, 042904 (2006).
\newline
42. E.L.Ince, Ordinary Differential Equations (Dover, New York,
1944).
\newline
43. E.G.Kalnins, W.Miller Jr and S.Post, Physics of Atomic Nuclei (to appear) 2009, arXiv:0908.4393
\newline
44.M.A.Rodriguez, P.Tempesta and P.Winternitz, Phys.Rev.E78 046608 (2008). 
\newline
45. M.A.Rodriguez, P.Tempesta and P.Winternitz,  J. Phys.: Conf. Ser. 175 012013 (2009). 

\newpage
\textbf{Figure captions}
\newline
Fig1.A trajectory for $V=\frac{\omega^{2}k_{1}^{2}}{18}(2b_{1} + 5x^{2} + \epsilon_{1} 4x\sqrt{b_{1}+x^{2}}) +\frac{\omega^{2} k_{2}^{2}}{18}(2b_{2} + 5y^{2} + \epsilon_{2} 4y\sqrt{b_{2}+y^{2}})$.
Parameter $\epsilon_{1}=1$, $\epsilon_{2}=1$, $\omega=3$, $k_{1}=1$,$k_{2}=3$,$b_{1}=3$, $b_{2}=5$, $v_{xo}=1$, $x_{o}=1$, $v_{yo}=-3$,
$y_{o}=1$, t=[0,20].
\newline
\newline
Fig2.A trajectory for $V=\frac{\omega^{2}k_{1}^{2}}{18}(2b_{1} + 5x^{2} + \epsilon_{1} 4x\sqrt{b_{1}+x^{2}}) +\frac{\omega^{2} k_{2}^{2}}{18}(2b_{2} + 5y^{2} + \epsilon_{2} 4y\sqrt{b_{2}+y^{2}})$.
Parameter $\epsilon_{1}=1$, $\epsilon_{2}=1$, $\omega=3$, $k_{1}=3$,$k_{2}=4$,$b_{1}=3$, $b_{2}=5$, $v_{xo}=1$, $x_{o}=1$, $v_{yo}=-3$,
$y_{o}=1$, t=[0,20].
\newline
\newline
Fig3.A trajectory for $V=\frac{\omega^{2}k_{1}^{2}}{18}(2b_{1} + 5x^{2} + \epsilon_{1} 4x\sqrt{b_{1}+x^{2}}) +\frac{\omega^{2} k_{2}^{2}}{18}(2b_{2} + 5y^{2} + \epsilon_{2} 4y\sqrt{b_{2}+y^{2}})+\frac{\omega^{2} k_{3}^{2}}{18}(2b_{3} + 5z^{2} + \epsilon_{3} 4z\sqrt{b_{3}+z^{2}})$. 
Parameter $\epsilon_{1}=1$, $\epsilon_{2}=1$, $\epsilon_{3}=1$,$\omega=3$, $k_{1}=7$, $k_{2}=11$, $k_{3}=4$, $b_{1}=3$, $b_{2}=5$, $b_{3}=7$, $v_{xo}=1$, $x_{o}=1$, $v_{yo}=-3$,
$y_{o}=1$, $z_{o}=1$, $v_{zo}=2$, t=[0,20].
\newline
\newline
Fig4.A trajectory for $V=\frac{\omega^{2} k_{1}^{2}}{18}(2b_{1} + 5x^{2} + \epsilon_{1} 4x\sqrt{b_{1}+x^{2}}) +\frac{\omega^{2} k_{2}^{2}}{18}(2b_{2} + 5y^{2} + \epsilon_{2} 4y\sqrt{b_{2}+y^{2}})+\frac{\omega^{2}k_{3}^{2}}{18}(2b_{3} + 5z^{2} + \epsilon_{3} 4z\sqrt{b_{3}+z^{2}})$. Parameter $\epsilon_{1}=1$, $\epsilon_{2}=1$, $\epsilon_{3}=1$, $\omega=3$, $k_{1}=5$, $k_{2}=6$, $k_{3}=2$ $b_{1}=3$, $b_{2}=5$, $b_{3}=7$, $v_{xo}=1$, $x_{o}=1$, $v_{yo}=-3$, $y_{o}=1$,  $z_{o}=1$, $v_{zo}=2$, t=[0,20].

%\end{flushleft}
\end{document}